\documentclass[a4paper]{jpconf}
\usepackage[T1]{fontenc} 
\usepackage{amsmath}
\usepackage{graphicx}
\usepackage{subcaption}
\usepackage{amssymb}
\usepackage{url}
\usepackage{float}
\usepackage{afterpage}
\newcommand{\myat}{{\fontfamily{pag}\selectfont @}}
\usepackage[labelfont=bf]{caption} 
\begin{document}
\nocite{*}
\title{Potential of the MeerKAT telescope to detect the stimulated decay of axion-like particles}
\author{A Ayad and G Beck}
\address{School of Physics, University of the Witwatersrand, Private Bag 3, WITS-2050, Johannesburg, South Africa}
\ead{ahmed\myat aims.edu.gh}
\begin{abstract}
A prominent aspect of the hunt for cold dark matter is looking for light scalar candidates such as axion-like particles (ALPs). The coupling between ALPs and photons allows for the spontaneous decay of ALPs into pairs of photons. It has been previously shown that stimulated ALP decay rates can become significant on cosmic time scales. Furthermore, it has been claimed, in several recent works, that ALPs can gravitationally thermalize and form macroscopic condensates. Consequently, the photon occupation number of ambient populations (like the cosmic microwave background) can receive Bose enhancement in dense ALP clumps and grows exponentially. For cold dark matter ALPs, this can lead to radio emissions produced from this process and could be observed by the forthcoming radio telescopes. In this work, we investigate the detectability of such a radio signature from some astrophysical targets using the MeerKAT radio telescopes. The results show that the MeerKAT telescope is able to probe the ALPs parameter space with limits reaching the current level of the CAST experiment and the potential level of the IAXO experiment with an arcminute visibility taper.
\end{abstract}

\section{Introduction} \label{sec.1}

Large amounts of stable cold dark matter are required to explain many astrophysical and cosmological observations \cite{komatsu2011seven}. These include the rotation curves of galaxies \cite{rubin1970rotation}, gravitational lensing by galaxy clusters \cite{blandford1992cosmological}, and the power spectrum of the cosmic microwave background \cite{fixsen1996cosmic}. However, the nature of this dark matter is still one of the most perplexing unsolved problems in cosmology and particle physics. This problem can potentially be explained by the presence of non-baryonic dark matter candidates such as axions and axion-like particles (ALPs). Axions \cite{peccei1977cp} are pseudo-scalar bosons arising in the Peccei-Quinn mechanism proposed to solve the problem of charge-parity invariance of the strong interactions in quantum chromodynamics (QCD) \cite{peccei2008strong}. While ALPs are similar particles often predicted by a variety of string-theoretic extensions to the standard model of particle physics \cite{anselm1982second}. Axions and ALPs have the same characteristics that are determined by their coupling with two photons. The main difference between them is that the coupling parameter for the QCD axions is directly related to the axion mass, however, this is not necessarily the case for the generic ALPs. These axions and ALPs are believed to be very light and weakly interacting with most of the standard model particles, see reference \cite{asztalos2006searches}. Because of these properties, it is suggested that the total content of the cold dark matter in the universe, or at least a portion of it, may well consist of axions and ALPs \cite{preskill1983cosmology, abbott1983cosmological, dine1983not}.

In recent years, the search for dark matter particles in the form of axions and ALPs has grown enormously, for simplicity we will use the term ALPs to refer to both axions and ALPs. Most of the current search for direct detection of ALPs depends essentially on their coupling with photons via a two-photon vertex \cite{raffelt1988mixing}. This coupling can allow for the conversion of ALPs into single photons $a \rightarrow \gamma$ through the Primakoff effect in the presence of an external magnetic field, see recent works \cite{ayad2020probing, ayad2019phenomenology}. This coupling may also allow for the decay of ALPs into pairs of photons $a \rightarrow \gamma + \gamma$, see recent works \cite{ayad2020quantifying, ayad2020potential}. While the first process received more attention in the research, we are now focusing on deepening our understanding of the second process. In our previous work \cite{ayad2020quantifying}, we studied the stimulated decay of ALPs and the influence of plasma effects on this process. Then we compared the predicted fluxes produced from this process with the sensitivity of the Square Kilometer Array (SKA) and the MeerKAT radio telescopes. We found that the point source sensitivity of the MeerKAT telescope had very limited potential for probing the ALP coupling-mass parameter space. In this work, we study the possibility of using different taper scales as an effective tool to improve the limits that the MeerKAT telescope can put on the coupling parameter between ALPs and photons. This tapering approach has been used for indirect radio detection of dark matter in a previous work \cite{regis2017dark}. We find that the use of taper calculations is able to improve the MeerKAT limits on the ALP-photon coupling by about an order of magnitude.

The structure of this paper is as follows. In Sections \ref{sec.2}, we briefly discuss the spontaneous and stimulated decay of ALPs into pairs of photons. Then, we discuss the sensitivity of the MeerKAT telescope and the use of the tapering approach to improving it in Section \ref{sec.3}. Next, in Section \ref{sec.4}, we check the capability of the MeerKAT telescope to put new limits on the coupling parameter between ALPs and photons. Finally, our conclusion is provided in Section \ref{sec.5}.

\section{{Spontaneous and stimulated decay of ALPs }} \label{sec.2}

The fundamental process that describes the phenomenology of ALPs is the interaction between them and photons through the ALP-two-photon vertex. The Lagrangian equation that describes the electromagnetic field in interaction with the ALP field is given by \cite{raffelt1988mixing}
\begin{equation} \label{eq.1}
\mathrm{\ell}_{\rm a\gamma} = - \frac{1}{4} g_{\rm a\gamma} \mathrm{F}_{\mu \nu} \tilde{\mathrm{F}}^{\mu \nu} a = g_{\rm a\gamma}\, \mathbf{E} \cdot \mathbf{B} \, a \:.
\end{equation}
Here $a$ is the ALP field, $\mathrm{F}_{\mu \nu}$ is the electromagnetic field strength, $\tilde{\mathrm{F}}^{\mu \nu}$ is its dual, and $g_{\rm a\gamma}$ is the ALP-photon coupling parameter. While $\mathbf{E}$ and $\mathbf{B}$ represent the electric and magnetic fields respectively. The presence of the ALP-two-photon interaction vertex makes possible the spontaneous decay of an ALP with mass $m_{\rm a}$ into pair of photons, each with a frequency $\nu=m_{\rm a} / 4 \pi$. The decay time of ALPs can be expressed as inverse of their decay rate in terms of the ALP mass and the ALP-photon coupling as in \cite{kelley2017radio}
\begin{equation} \label{eq.2}
\tau_{\rm a} \equiv \Gamma_{\rm pert}^{\rm -1} = \frac{64 \pi}{m_{\rm a}^{\rm 3} \; g_{\rm a\gamma}^{\rm 2}} \:.
\end{equation}
For the typical QCD axion with mass $m_{\rm a} \sim 10^{\rm -6} \; {\rm eV}$ and coupling with photons $g_{\rm a\gamma} \sim 10^{\rm -12} \; {\rm GeV}^{\rm -1}$, the lifetime $\tau_{\rm a}$ is about $1.32 \times 10^{\rm 47} \; {\rm s}$. This lifetime for axions is large enough comparing to the current age of the universe $4.3 \times 10^{\rm 17} \; {\rm s}$ to say that this type of particle is very stable on the cosmological scale. This might be the main reason for not focusing on searching for a detectable signal produced from the spontaneous decay of ALPs in the literature. However, this is ignoring the fact that ALPs are bosons. They have very low mass and accordingly they may exist with very high occupation numbers forming a Bose-Einstein condensate with only short-range order. These condensates may thermalize due to their gravitational attraction and self-interactions to spatially localized clumps comprising the dark matter halos around galaxies \cite{sikivie2009bose, hertzberg2018scalar}. In the presence of background photon population with high occupation number $f_{\rm \gamma}$, the stimulated decay of ALPs condensate is highly likely with an effective decay rate expressed as
\begin{equation} \label{eq.3}
\Gamma_{\rm eff}= \Gamma_{\rm pert}(1+2 f_{\rm \gamma}) \:.
\end{equation}
Such systems with very high occupation numbers are well described by a classical field approximation \cite{guth2015dark}. Using the classical equation of motion; the stimulated decay time of ALPs was calculated to be just about $10^{\rm -7} \; {\rm s}$, which is dramatically small comparing to the spontaneous decay time. This is problematic because theoretically ALPs are expected to be left-over from the early universe and their very short lifetime implies that they must decay so rapidly. Therefore according to this scenario, ALPs must have vanished a long time ago, and we can not hope for any current observable signal to remain due to their stimulated decay. The expansion of the universe and the plasma effects slow down the decay rate of ALPs and explain the huge discrepancy between the classical and the standard calculations \cite{alonso2020wondrous}. Considering that the mass range $m_{\rm 0}\text{--}m_{\rm 1}$ (these masses were placeholders for the range allowed by the cosmic plasma) will be able to sustain large populations into the present epoch while undergoing stimulated decay \cite{madau2014cosmic}, we estimate the radio emissions which can be produced from the stimulated decay of ALPs if they form the total dark matter content of galactic halos. Then, we compare the predicted fluxes with the sensitivity of the MeerKAT telescope in order to determine the region of the mass-coupling parameter space that is accessible to MeerKAT. 

\section{The MeerKAT radio telescope and the taper sensitivity calculations} \label{sec.3}

The MeerKAT radio telescope is a precursor for the SKA system in South Africa which would be the most sensitive radio telescope ever upon its completion \cite{booth2012overview}. This project has the potential to produce a high-performing contribution to obtain new cosmological constraints from a large sky survey. This makes it the most aspirant radio telescope to detect the possible emissions produced from the stimulated decay of ALPs and probe their parameter space. 

In our analysis, we determine the MeerKAT sensitivities via the Stimela \cite{makhathini2018advanced} software package. This allows us to find the rms image noise after simulating MeerKAT observations with the NVSS source catalog as a sky model \cite{condon1998nrao}. To do this, Stimela makes use of CASA \cite{mcmullin2007casa} for observation simulation, RFIMasker\footnote{\url{https://github.com/bennahugo/RFIMasker}} to mask out frequency channels with static interference, Meqtrees \cite{noordam2010meqtrees} for calibration and visibility simulation, and WSClean \cite{offringa2014wsclean} for imaging (at robust weighting $1$). All our sensitivities apply to the MeerKAT L-band (890 MHz to 1.65 GHz) and assumed a channel width of 1 MHz for the simulation. To improve the MeerKAT sensitivity we employ different taper scales on the visibilities. The use of a taper has been reported in \cite{regis2017dark} in ATCA observations of dwarf galaxies to probe WIMP dark matter. The calculations of the effect of tapering are drawn from \cite{beck2021} and make use of the Stimela package \cite{makhathini2018advanced} to realistically simulate MeerKAT sensitivities. A UV taper acts to down-weight the contribution of long baselines to interferometric observations. The consequence of which is the reduction in the contribution of small-scale sources \cite{wilson2011techniques}. This is of great use in hunting for ALP emissions as they should be diffuse, emanating from an extended halo around the center of observed cosmic structures. Here we will examine the effect of the chosen taper scale on MeerKAT's sensitivity to ALP emissions.

\section{Results and discussion} \label{sec.4}

\begin{figure}[t!]
\begin{subfigure}{0.48\textwidth}
\centering
\includegraphics[width=\linewidth]{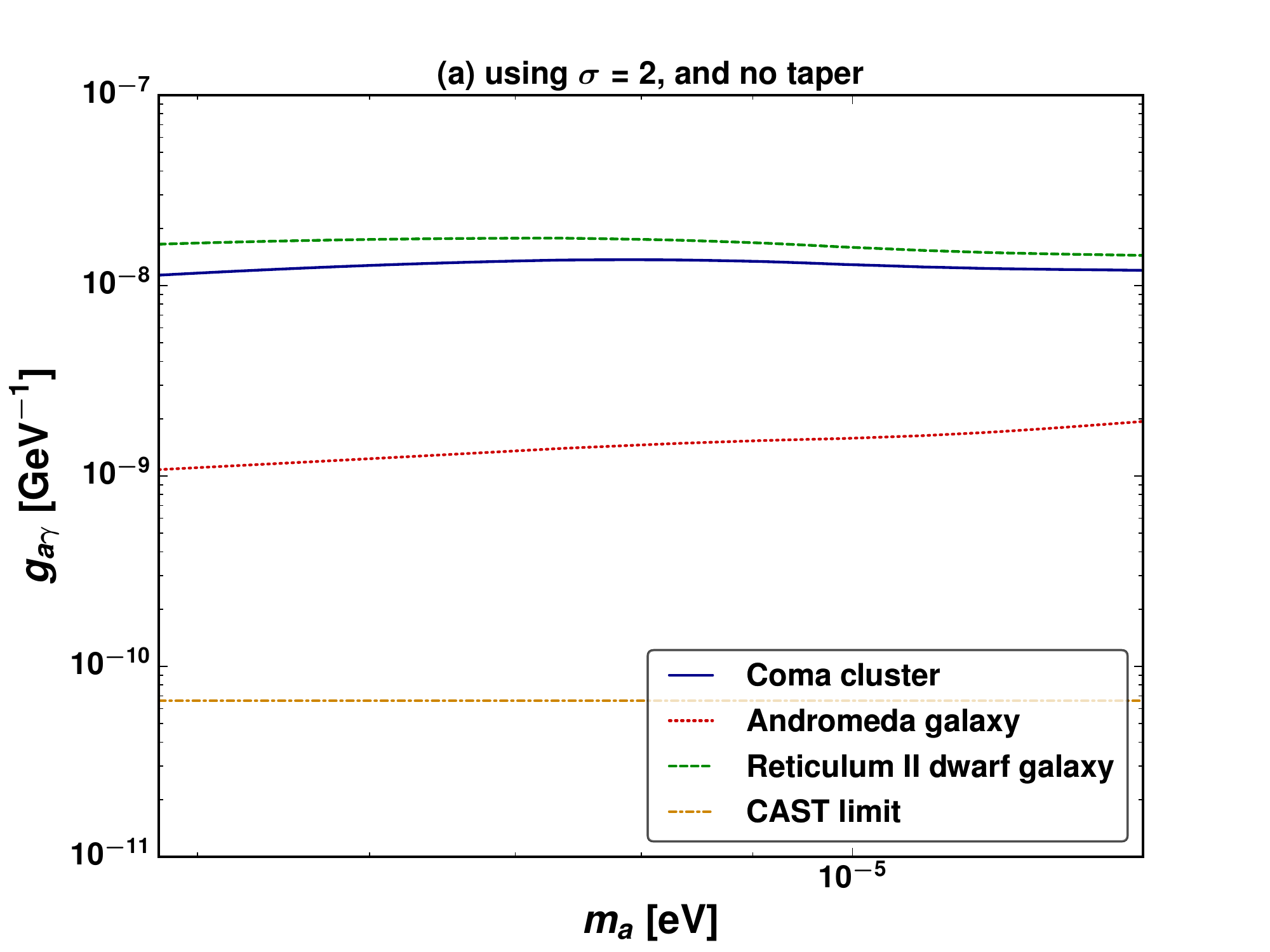}
\end{subfigure}\hspace*{\fill}
\begin{subfigure}{0.48\textwidth}
\centering
\includegraphics[width=\linewidth]{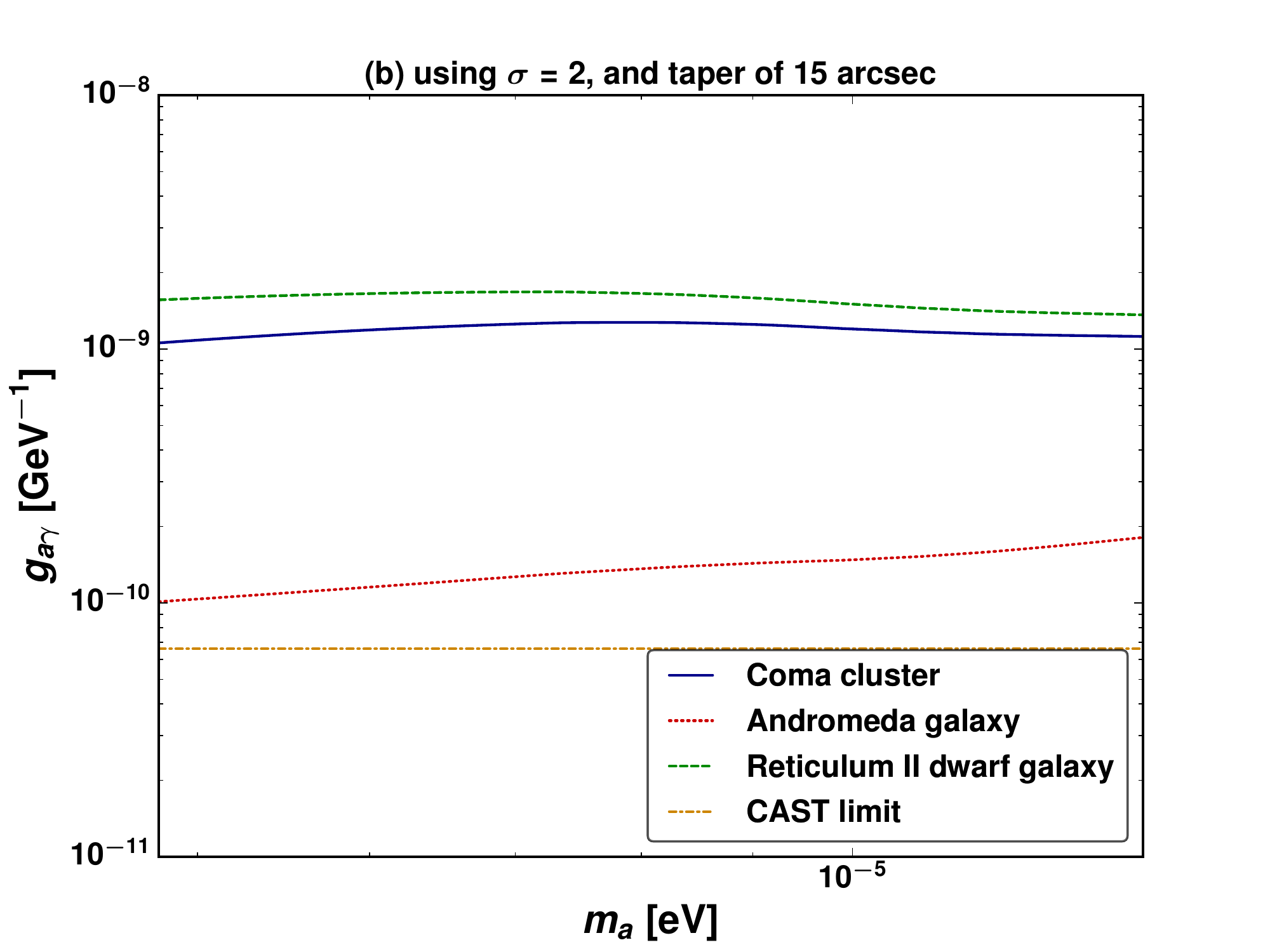}
\end{subfigure}

\medskip
\begin{subfigure}{0.48\textwidth}
\centering
\includegraphics[width=\linewidth]{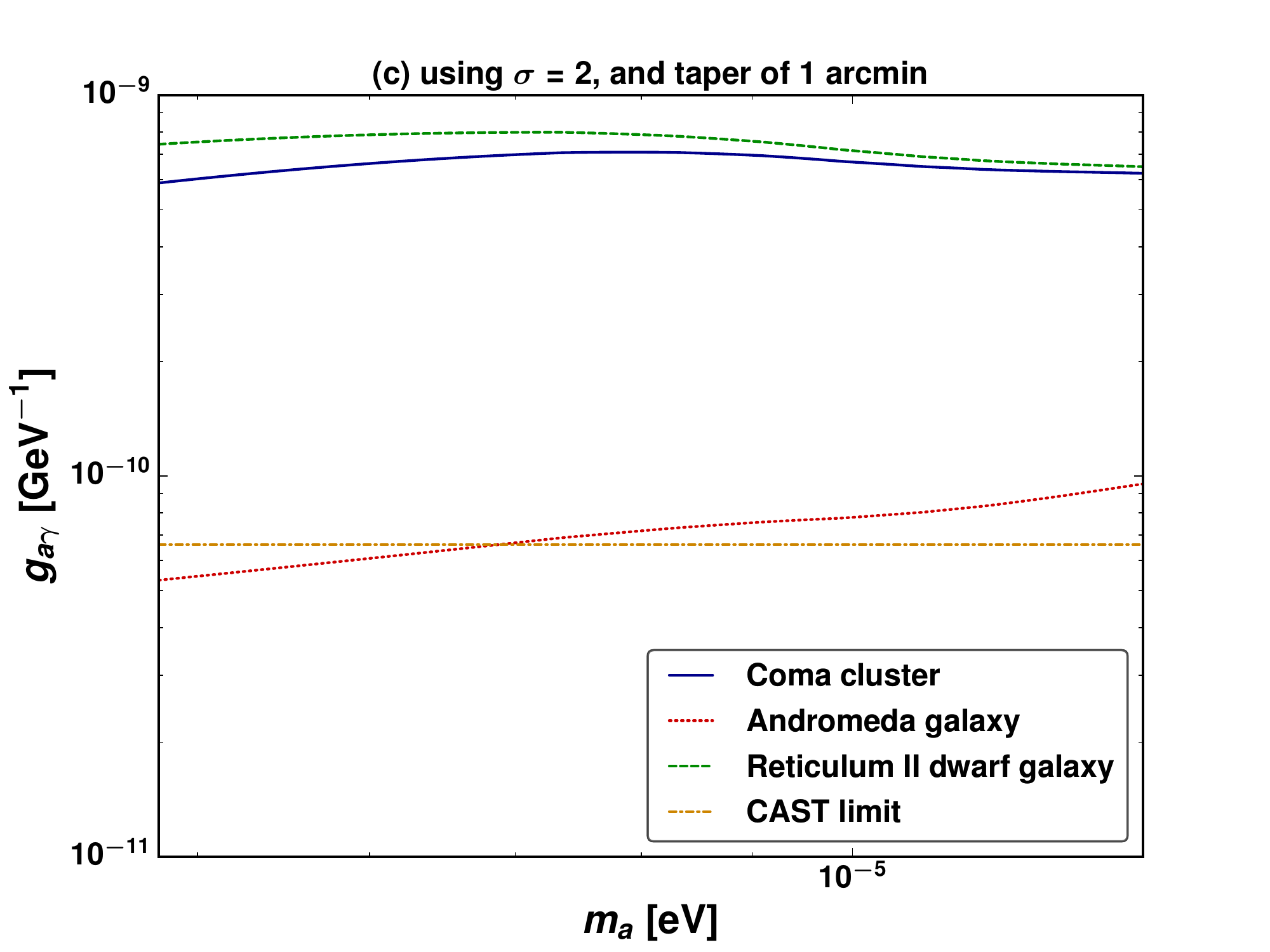}
\end{subfigure}\hspace*{\fill}
\begin{subfigure}{0.48\textwidth}
\centering
\includegraphics[width=\linewidth]{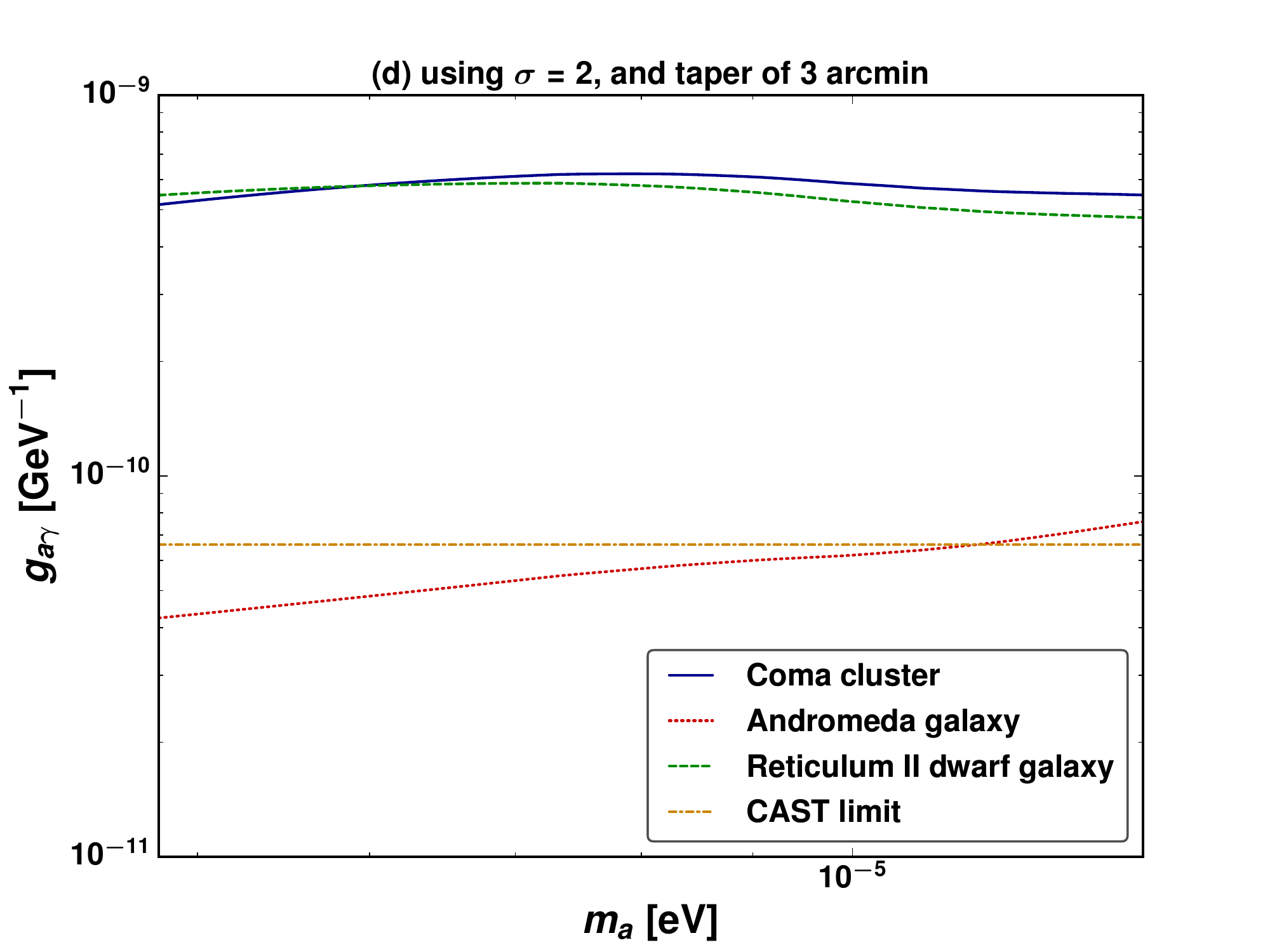}
\end{subfigure}

\medskip
\begin{subfigure}{0.48\textwidth}
\centering
\includegraphics[width=\linewidth]{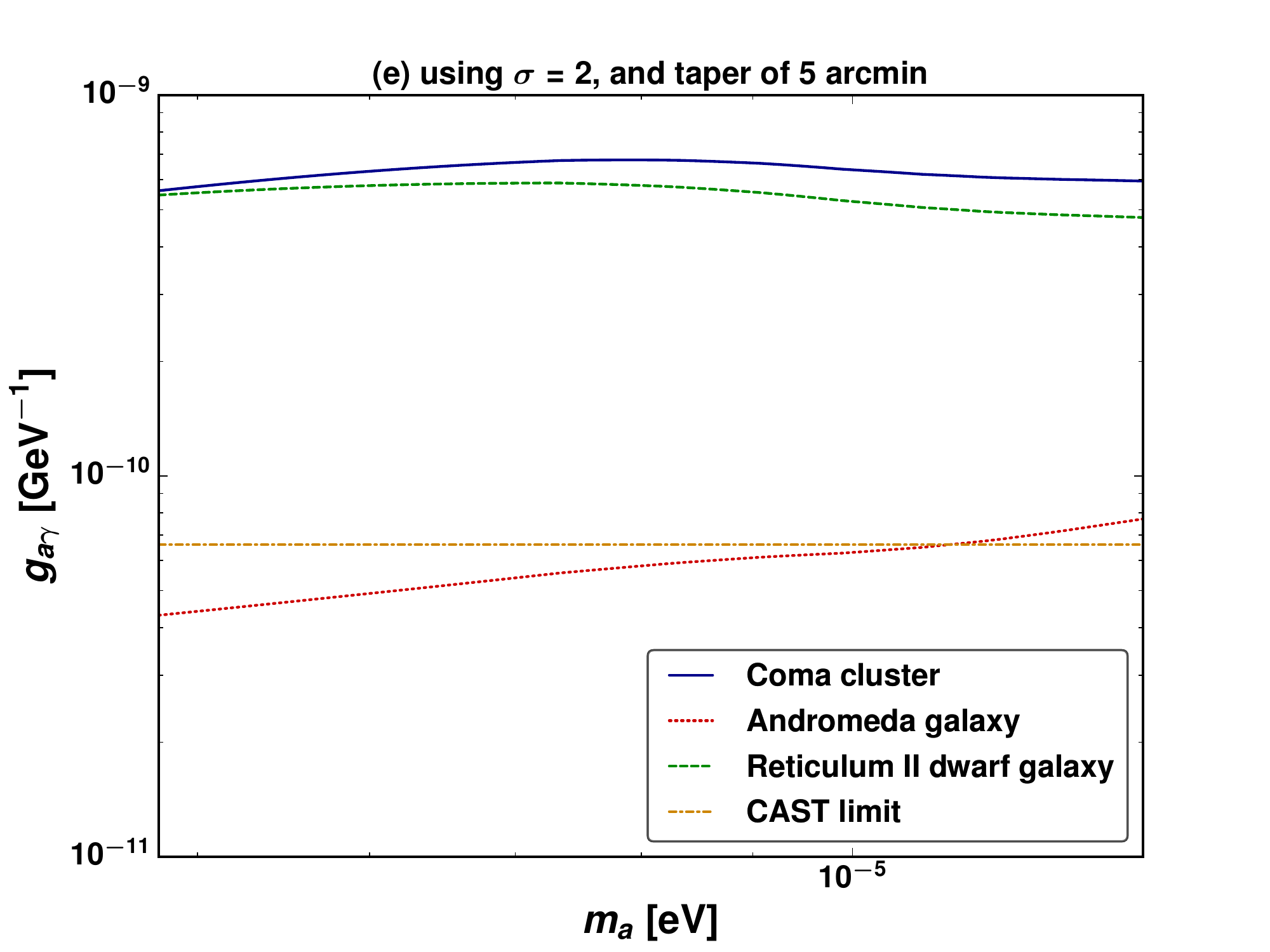}
\end{subfigure}\hspace*{\fill}
\begin{subfigure}{0.48\textwidth}
\centering
\includegraphics[width=\linewidth]{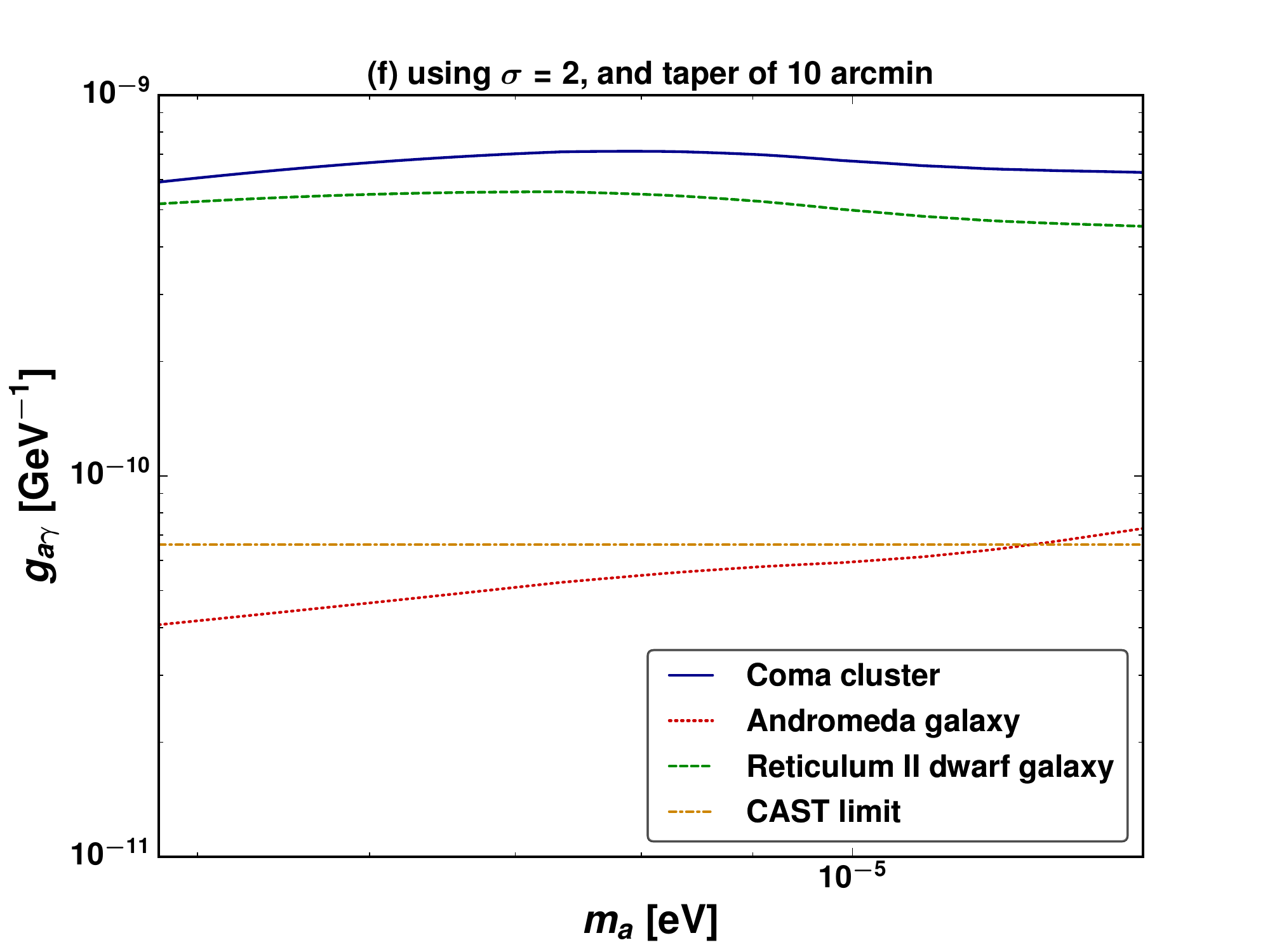}
\end{subfigure}

\caption{The potential non-observation constraints with the MeerKAT telescopes for 2$\sigma$ confidence interval using halo surface brightness from ALP decay. In this plot, taper with different scales is employed on the visibilities. The CAST limit is taken from \cite{anastassopoulos2017new}.}
\label{fig.1}
\end{figure}

The stimulated decay of ALPs would lead to an exponential growth in the photon occupation number, and therefore we expect to produce a large flux of photons with a frequency of about half the ALP mass. Hence, we hunt for a signal that would appear as a narrow spectral line broadened by the ALPs velocity dispersion. The radio emissions per unit surface area can be given by the following expression
\begin{equation} \label{eq.6.7.4}
S_{\rm SB} = \frac{\mathrm{\Gamma}_{\rm eff}}{4 \pi} \frac{\Delta \nu}{\nu} \int_{\rm R}^{\rm \infty} \frac{\rho_{\rm a}(r)}{\sqrt{r^{\rm 2}-R^{\rm 2}}} \ dr \:.
\end{equation}
Here the factor $\Delta \nu/\nu$ describes the effect of line broadening, $\rho_{\rm a}$ is the density of the dark matter halo, $R$ is the radius at which we determine the surface brightness. In our analysis, in order to determine the dark matter density in the galaxy halos, we use the NFW profile for the Coma cluster \cite{lokas2003dark} as well as for the Andromeda galaxy \cite{tamm2012stellar} and we use the Einasto profile for the Reticulum II dwarf galaxy \cite{regis2017dark}. The MeerKAT is currently equipped with receivers that are sensitive to cover a frequency range from 890 MHz to 1.65 GHz \cite{braun2019anticipated}. This range of frequencies corresponds to photons produced from the decay of ALPs in the mass range $4.80 \times 10^{\rm -6} \text{--} 1.36 \times 10^{\rm -5} \ {\rm eV}$. With this range of ALPs masses, we find that the radio emissions per unit surface are quite difficult to detect considering the point-source sensitivity of the MeerKAT telescope\footnote{For the MeerKAT sensitivity calculations, see the online sensitivity calculator at: \url{https://skaafrica.atlassian.net/wiki/spaces/ESDKB/pages/41091107/Sensitivity+calculators.}}. In panel (a) of Figure~\ref{fig.1}, we illustrate the potential non-observation constraints with the MeerKAT telescope for 2$\sigma$ confidence interval using point source sensitivities. This point source sensitivity produces weak limits on the values of the coupling strength $g_{\rm a\gamma} \sim 1.09 \times 10^{\rm -9} \ {\rm GeV}^{\rm -1}$, comparing to the current limits put by the CAST experiment, i.e. $g_{\rm a \gamma} \lesssim 6.6 \times 10^{\rm -11} \ {\rm GeV}^{\rm -1}$ \cite{anastassopoulos2017new}, and the potential limits of the IAXO experiment, i.e. $g_{\rm a \gamma} \lesssim {\rm few} \times 10^{\rm -11} \ {\rm GeV}^{\rm -1}$ \cite{armengaud2019physics}. In panels (b) to (f) of Figure~\ref{fig.1}, we improve these results using tapering effects with different scales on the visibilities on the MeerKAT sensitivity with 2$\sigma$ confidence interval. The results showing the taper effect are summarized in Table~\ref{tab.1} using 2$\sigma$ confidence interval as well as 5$\sigma$ confidence interval. We remark that the use of the 2$\sigma$ confidence interval allows us to estimate the exclusion limits in the event of non-observation, whereas 5$\sigma$ confidence interval demonstrates the discovery potential. From the table, we find that the benefits of a taper saturate around scales of 1 arcminute, but provide a sensitivity gain of around an order of magnitude. The projected limits now are stronger as about $g_{\rm a\gamma} \sim 4.09 \times 10^{\rm -11} \ {\rm GeV}^{\rm -1}$ in the most promising case, and in general overlapping now with the limits of the IAXO and CAST experiments.

It is worth noting here that future works might address more reasons responsible for boosting these radio emissions and that the MeerKAT telescope will also receive an upgrade of 20 additional dishes by 2023 which will enhance its sensitivity. Hence, the MeerKAT telescope will be able to play a more important role in parallel with other experiments such as CAST and IAXO experiments to rule out new regions of the parameter space of ALPs.
\begin{center}
\begin{table}[t!]
\caption{\label{tab.1} Summarizing the limits of the minimum ALP-photon coupling that can be probed by the MeerKAT radio telescope with the $2\sigma$ and $5\sigma$ confidence intervals within the MeerKAT sensitivity range for the ALP mass and using tapered sensitivities with different scales.}
\centering
\begin{tabular}{@{}*{7}{l}}
\br
Taper scale & Minimum coupling $g_{\rm min}$ with $\sigma=2$ & Minimum coupling $g_{\rm min}$ with $\sigma=5$ \\
\mr
\verb"untapered" & $1.09 \times 10^{\rm -9}$ GeV$^{\rm -1}$ & $1.71 \times 10^{\rm -9}$ GeV$^{\rm -1}$ \\
\verb"15 arcsec" & $1.01 \times 10^{\rm -10}$ GeV$^{\rm -1}$ & $1.60 \times 10^{\rm -10}$ GeV$^{\rm -1}$ \\
\verb"1 arcmin" & $5.33 \times 10^{\rm -11}$ GeV$^{\rm -1}$ & $8.42 \times 10^{\rm -11}$ GeV$^{\rm -1}$\\
\verb"3 arcmin" & $4.24 \times 10^{\rm -11}$ GeV$^{\rm -1}$ & $6.70 \times 10^{\rm -11}$ GeV$^{\rm -1}$\\
\verb"5 arcmin" & $4.31 \times 10^{\rm -11}$ GeV$^{\rm -1}$ & $6.81 \times 10^{\rm -11}$ GeV$^{\rm -1}$\\
\verb"10 arcmin" & $4.07 \times 10^{\rm -11}$ GeV$^{\rm -1}$ & $6.43 \times 10^{\rm -11}$ GeV$^{\rm -1}$\\
\br
\end{tabular}
\end{table}
\end{center}

\section{Conclusion} \label{sec.5}

Axion-like particles can form the total abundance of the cold dark matter in the universe. In this work, we illustrated that the MeerKAT radio telescope will be capable of probing the parameter space of ALPs by hunting for the radio emissions produced from their stimulated decay. The potential non-observation limits from this study are superior to the current limits put by the CAST experiment and comparable to the expected limits that can be reached by the IAXO experiment in future. In particular, the MeerKAT sensitivity can reach lower limits of the ALP-photon coupling as strong as $4.07 \times 10^{\rm -11}$ GeV$^{\rm -1}$. To obtain these results we had to use a visibility taper to improve the limits on the minimum detectable coupling. The taper provides an improvement on the limits by around an order of magnitude across the studied mass ranges. Indeed, increasing the sensitivity reach of the next-generation radio telescopes such as the MeerKAT telescope on probing the ALP-photon coupling will allow it to play a strong complementary role to experiments like CAST and IAXO in exploring the ALP parameter space. This undoubtedly should have a lot of influence in our search for direct evidence of the presence of cold dark matter in the form of ALPs using the next-generation of radio telescopes which are expected to receive more attention within the coming years.
\section*{Acknowledgments}

This work is based on the research supported by the South African Research Chairs Initiative of the Department of Science and Technology and National Research Foundation of South Africa (Grant No 77948). A. Ayad acknowledges support from the Department of Science and Innovation/National Research Foundation (DSI/NRF) Square Kilometre Array (SKA) post-graduate bursary initiative under the same Grant. G. Beck acknowledges support from a National Research Foundation of South Africa Thuthuka grant no. 117969. The authors would like also to offer special thanks to Prof. S. Colafrancesco, who, although no longer with us, continues to inspire by his example and dedication to the students he served over the course of his career. 

\section*{References}
\bibliography{references}
\end{document}